# Wireless Communications with Programmable Metasurface: Transceiver Design and Experimental Results

Wankai Tang [1], Xiang Li [1], Jun Yan Dai [2], Shi Jin [1], Yong Zeng [3], Qiang Cheng [2], Tie Jun Cui[2]

[1] National Mobile Communications Research Laboratory, Southeast University, Nanjing 210096, P. R. China.
[2] State Key Laboratory of Millimeter Waves, Southeast University, Nanjing, 210096, P. R. China.
[3] School of Electrical and Information Engineering, The University of Sydney, Sydney, NSW 2006, Australia.

**Abstract:** Metasurfaces have drawn significant attentions due to their superior capability in tailoring electromagnetic waves with a wide frequency range, from microwave to visible light. Recently, programmable metasurfaces have demonstrated the ability of manipulating the amplitude or phase of electromagnetic waves in a programmable manner in real time, which renders them especially appealing in the applications of wireless communications. To practically demonstrate the feasibility of programmable metasurfaces in future communication systems, in this paper, we design and realize a novel metasurface-based wireless communication system. By exploiting the dynamically controllable property of programmable metasurface, we firstly introduce the fundamental principle of the metasurface-based wireless communication system design. We then present the design, implementation and experimental evaluation of the proposed metasurface-based wireless communication system with a prototype, which realizes single carrier quadrature phase shift keying (QPSK) transmission over the air. In the developed prototype, the phase of the reflected electromagnetic wave of programmable metasurface is directly manipulated in real time according to the baseband control signal, which achieves 2.048 Mbps data transfer rate with video streaming transmission over the air. Experimental result is provided to compare the performance of the proposed metasurface-based architecture against the conventional one. With the slight increase of the transmit power by 5 dB, the same bit error rate (BER) performance can be achieved as the conventional system in the absence of channel coding. Such a result is encouraging considering that the metasurface-based system has the advantages of low hardware cost and simple structure, thus leading to a promising new architecture for wireless communications.

## I. Introduction

Since the introduction of the first generation analog wireless communications system in the 1980s, the mobile access technology has undergone revolutionary advancement for about every decade. Now the fifth generation of cellular mobile communication (5G) system is being developed rapidly, which is anticipated to support a diverse variety of use cases including enhanced mobile broadband (eMBB), ultra-reliable low latency communications (URLLC) and massive machine type communications (mMTC) [1]. The wide-scale commercial deployment of 5G is anticipated to begin around 2020 and the technical research on beyond 5G (B5G) or 6G has already been initiated by several groups [2]. Among the various new technologies that have been proposed, the introduction of artificial intelligence [3], the use of the Terahertz band [4] and the integration of space and terrestrial networks [5] are regarded as the most promising key technologies in the future B5G era. Besides, a new type of electromagnetic surface structure called *programmable metasurface*, which may fundamentally change the basic hardware structure of wireless communication system, is emerging and has attracted attentions [6] [7]. The past few years have witnessed the great success of metasurface in a wide range of applications including imaging [8], antenna [9], radar [10] and hologram [11], thanks to its great flexibility in manipulating electromagnetic waves [12]. However, the application of metasurface in wireless communications is still in its infancy. A recent study has shown that programmable metasurface can be used to proactively improve the channel environment for wireless communications [13].

On the other hand, the traditional quadrature sampling transceiver using zero intermediate frequency (zero-IF) or superheterodyne structure has been applied in mobile communication for many years with great success [14] [15]. However, such conventional architecture also faces many critical challenges in wireless communication systems that require ultra wide bandwidth, high processing capability and low power consumption in the future. The digitization of wireless communication systems has evolved from baseband to radio frequency (RF) chains and antennas, which calls for an imperative need to develop more flexible hardware architecture for future wireless communication systems. For instance, the direct antenna modulation technique has been previously proposed to directly generate modulated RF signals using time-varying antennas, but it has the limitation of only supporting several inefficient basic modulation schemes such as on-off keying (OOK) [16] and frequency shift keying (FSK) [17] [18].

In this paper, we exploit the advanced programmable metasurface to realize a novel wireless communication system in which the hardware structure of the transmitter is based on programmable metasurface entirely. Phase modulation of the reflected electromagnetic wave can be achieved directly by electrically controlling the reflection coefficient of our metasurface-based transmitter, which has the advantages of low hardware cost, low energy consumption and simple structure. We implement a metasurface-based QPSK prototype system and conduct several over-the-air transmission experiments to verify the feasibility, reliability, and stability of this new transmitter architecture. The experimental results prove that this novel system can work reliably and stably.

The rest of the paper is organized as follows. Section II provides a brief overview of programmable metasurface. Section III describes the design principle and process of our proposed metasurface-based wireless communication system, including fundamental principle, transmitter design, frame structure and receiver design. Section IV shows our prototype setup. Section V presents some experimental results for over-the-air transmissions. Section VI outlines several challenges and opportunities for further research. Lastly, we conclude the paper in VII.

## II. CONCEPT OF PROGRAMMABLE METASURFACE

Metasurfaces are two-dimensional artificial subwavelength structures with unique electromagnetic properties such as negative refraction, which are usually cannot be found in nature [19]. They are typically designed by deliberately arranging a set of well-designed sophisticated small scatterers or apertures in a regular array to achieve the desired ability for guiding and controlling the flow of electromagnetic waves [20], which can be dragged into almost any desired configuration [21].

Conventionally, the desired ability and configuration for manipulating electromagnetic waves of metasurfaces are fixed. For example, the reflection/transmission coefficient of conventional metasurfaces is constant, which means that the phase/magnitude profiles are fixed once the metasurfaces are fabricated [11]. Such early generation of metasurfaces is regarded as "analog metasurfaces" since their control characteristics for electromagnetic waves cannot be dynamically adjusted, which are typically described by effective medium parameters [12].

Recently, the invention of programmable metasurfaces with reconfigurable electromagnetic parameters offers an effective way to overcome the above shortcoming of conventional metasurfaces [12] [22]. Programmable metasurfaces can dynamically change and manipulate the amplitude, phase, polarization, and even orbital angular momentum of reflected/transmitted electromagnetic waves over their surface. This makes programmable metasurfaces especially appealing for wireless communication systems.

In this paper, we explore and verify the feasibility of applying metasurface in wireless communication by using reflection-type programmable metasurface, which is composed of a number of unit cells to form an array surface.

Fig. 1(a) illustrates the top view schematic of the unit cell and its simplified equivalent circuit model. Four metallic rectangular patches, each pair of which is bridged by a varactor diode, constitute a unit cell. The varactor diodes are biased through via holes by the feeding network, which is composed of slotted copper plate at the bottom of the substrate. The unit cell can be modeled as a parallel resonant tank as shown in Fig. 1(a). The $C$, $R$, $L_1$, $L_2$, $Z_l$, $Z_0$ and $\Gamma$ represent the equivalent capacitance, resistance, inductance on the top of unit cell, the equivalent inductance at the bottom of unit cell, the equivalent load impedance of unit cell, the characteristic impedance of the air and the reflection coefficient of the unit cell, respectively. The capacitance of the simplified equivalent circuit model for unit cell is dominated by the varactor diode, which indicates that the load impedance can be tuned by the biasing voltage of varactor diode as $Z_l$ given by (1).

$$Z_l = \frac{jwL_2\left(jwL_1 + \frac{1}{jwC} + R\right)}{jwL_2 + \left(jwL_1 + \frac{1}{jwC} + R\right)}. \quad (1)$$

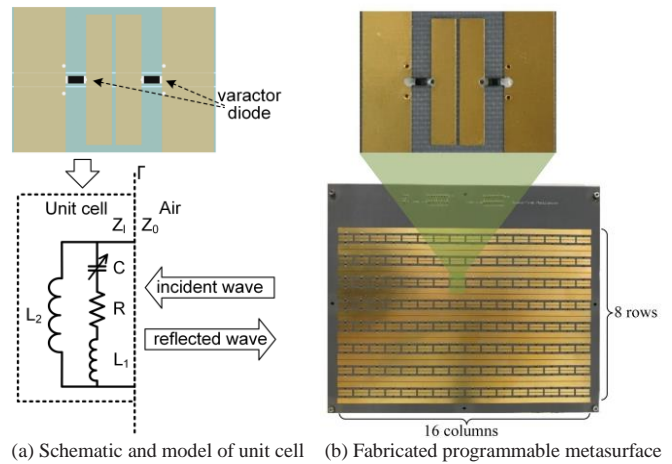

(a) Schematic and model of unit cell  (b) Fabricated programmable metasurface

Fig. 1. (a) Top view schematic of the unit cell and its simplified equivalent circuit model. (b) Fabricated sample of our programmable metasurface with a zoomed-in view of the unit cell.

The reflection coefficient is a parameter that describes the fraction of the electromagnetic wave reflected by an impedance discontinuity in the transmission medium. The reflection coefficient of the unit cell is determined by its equivalent load impedance $Z_l$ and the impedance towards the source $Z_0$, i.e., the characteristic impedance of the air in the considered system. Then the reflection coefficient of the unit cell can be written as [23]

$$\Gamma = \frac{Z_l - Z_0}{Z_l + Z_0}. \quad (2)$$

By combining (1) and (2), the phase of the reflection coefficient can be obtained. (3) illustrates the reflection phase tuning principle of the programmable metasurface in this paper.

$$\varphi(\Gamma) = arctan\left(\frac{Im(\Gamma)}{Re(\Gamma)}\right). \quad (3)$$

Fig. 1(b) shows the fabricated sample of our programmable metasurface with a zoomed-in view of the unit cell. A total of 8×16 unit cells are arranged periodically on the top of the substrate of the metasurface with an area of 176 ×252.8 mm². The junction capacitance of the varactor diode in each unit cell is affected by the biasing voltage, which works as the control signal that adjusts the capacitance value of the varactor diode, thus dynamically controls the reflection coefficient of each unit cell. By this way, the reflection coefficient of the whole metasurface is dynamically adjustable, making it "programmable".

In the following sections, we will present the design methodology for our proposed metasurface-based wireless communication systems, as well as the realization of the single-carrier QPSK transmission over the air using the fabricated programmable metasurface introduced above.

## III. METASURFACE-BASED WIRELESS COMMUNICATION SYSTEM

This section presents the architecture of the proposed metasurface-based wireless communication system in detail, including the fundamental principle, transmitter design, frame structure design and receiver design.

### A. Fundamental Principle

Fig. 2 shows the block diagram of the programmable metasurface. As shown in the figure, the incident wave, which is denoted as $E_i(t)$, impinges on the programmable metasurface from a feed antenna. The incident wave is a single-tone carrier signal and plays the role of carrier signal in our metasurface-based communication system. The programmable reflection coefficient of the metasurface, which essentially performs modulation, is expressed as $\Gamma(t)$. By definition, the reflection coefficient is the ratio of the complex amplitude of the reflected wave to that of the incident wave [24]. Thus, the reflected wave $E_r(t)$, which contains the modulated information at the frequency of the carrier signal (incident wave), can be expressed as

$$E_r(t) = E_i(t) \cdot \Gamma(t), \quad (4)$$

where $E_i(t) = A\cos(2\pi f_c t + \varphi_0)$. $A$, $f_c$, $\varphi_0$ represents the amplitude, the frequency and the initial phase of the incident single-tone electromagnetic wave, respectively. The Fourier transform of (4) is therefore expressed as

$$E_r(f) = A\left(\frac{e^{-j\varphi_0}\delta(f+f_c) + e^{j\varphi_0}\delta(f-f_c)}{2}\right) * \Gamma(f)$$
$$= A\left(\frac{e^{-j\varphi_0}\Gamma(f+f_c) + e^{j\varphi_0}\Gamma(f-f_c)}{2}\right), \quad (5)$$

where ∗ represents the convolution operation and $\delta(f)$ is the Dirac delta function. As can be observed from (5), the spectrum of the reflected wave $E_r(f)$ has been shifted to the vicinity of the carrier frequency $f_c$, and its shape is bounded by the spectrum of programmable reflectivity $\Gamma(f)$. This is in accordance with the up-conversion in the conventional wireless communication systems, though in the latter it is achieved by mixers. Such an up-conversion mechanism eliminates the need of mixers and filters. Instead, it is based on passive programmable metasurface with the advantages of low complexity, low cost, low power consumption and low heat dissipation.

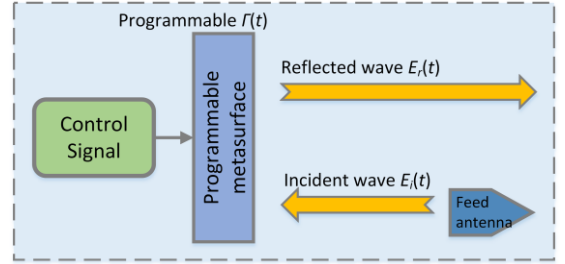

Fig. 2. The block diagram of the programmable metasurface.

In such metasurface-based architecture, the source data is carried by the reflection coefficient $\Gamma(t)$. Therefore, when designing a metasurface-based communication system, the key lies in how to design a mechanism for controlling the reflection coefficient $\Gamma(t)$ based on the source data and the desired modulation method. According to the reflection characteristics of the specific programmable metasurface, the desired mapping method can be well designed. In the following, we will illustrate how to design the programmable reflectivity $\Gamma(t)$ to construct a single-carrier QPSK wireless communication system.

### B. Transmitter Design

Fig. 3 shows the block diagram of a wireless communication system by using a reflectivity-programmable metasurface at the transmitter. The transmitter is completely based on programmable metasurface as shown in Fig. 3(a). The source bits to be transmitted is mapped to the reflection coefficient control signal of the metasurface to realize the modulation and emission of the reflected wave. In this paper, single-carrier QPSK modulation is implemented at the transmitter, for which the phase of the reflected wave is modulated into four different states. Therefore, the design of the metasurface reflection coefficient can be written as

$$\Gamma(t) = \sum_{n=1}^{N} \Gamma_n h(t - nT), \Gamma_n \in \{P_1, P_2, P_3, P_4\}, \quad (6)$$

where $\Gamma_n$ is the complex reflection coefficient corresponding to the $n^{th}$ message symbol that has four possible values. $T$ represents the symbol duration and $h(t)$ is the sampling function. $P_1$, $P_2$, $P_3$ and $P_4$ represent the complex values of the four different reflection coefficients. In the standard QPSK modulation, all data symbols have identical amplitude, but with 90 degrees phase difference between adjacent symbols. Furthermore, each symbol $\Gamma_n$ constitutes $\log_2 4 = 2$ information bits. For instance, $P_1$ represents '00', $P_2$ represents '01', $P_3$ represents '11' and $P_4$ represents '10' as shown in Table I. If the message to be transmitted is '00100111', the complex reflection coefficient should be sequentially set to '$P_1P_4P_2P_3$' during four consecutive message symbols by setting a corresponding control signal sequence for the metasurface.

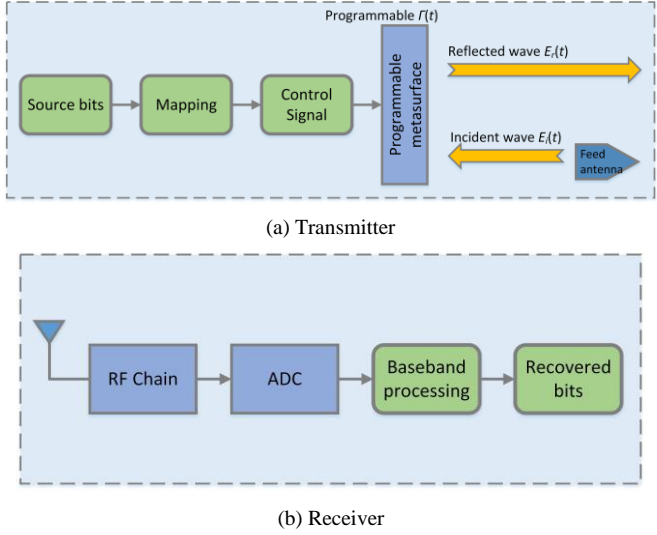

(a) Transmitter

(b) Receiver

Fig. 3. The block diagram of the proposed metasurface-based wireless communication system. (a) Transmitter. (b) Receiver.

TABLE I
MAPPING BETWEEN REFLECTION COEFFICIENT AND TRANSMISSION BITS

| Reflection coefficient | $P_1$ | $P_2$ | $P_3$ | $P_4$ |
|---|---|---|---|---|
| Transmission bits | 00 | 01 | 11 | 10 |

The metasurface used in this paper contains a total of 128 unit cells. In principle, the reflection characteristics of each unit cell can be controlled independently by their respective control signals to generate complex reflected electromagnetic waves, which thus enables multiple beams and has the great potential to be applied with MIMO technology. However, for ease of exposition and as the initial attempt to verify the concept of wireless communication with programmable metasurface, in this paper we focus on the identical control signal to control the reflection coefficients of all 128 unit cells. In this case, the modulations of the reflected waves on all units are the same, i.e., the same QPSK modulation is implemented for all reflected waves from every unit cell. The control circuit behind the programmable metasurface distributes the same control voltage to all 128 unit cells and amplifies the voltage signal to the voltage range required by the varactor diode of each unit cell.

The carrier frequency, or the frequency of the incident single-tone carrier signal is 4 GHz, at which the metasurface we used has the highest energy reflection efficiency (about 100% based on simulation results). The carrier frequency can be extended to the millimeter-wave band or terahertz band in the future by redesigning the corresponding programmable metasurface. Based on the measurement results conducted in the microwave anechoic chamber, the relationship between the control voltage of programmable metasurface and the phase of reflected wave at 4 GHz is shown in Fig. 4. It is observed that the control voltage and the reflected wave phase have a non-linear relationship, specifically, as the control voltage becomes larger, the varactor diode gradually reaches the saturation region, where its capacitance value remains almost constant. Therefore, the phase value of the reflected wave also tends to be unchanged under a large voltage.

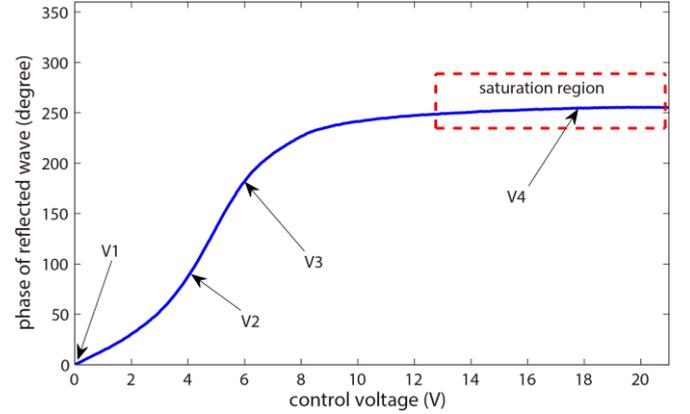

Fig. 4. The relationship between the control voltage of programmable metasurface and the phase of reflected wave.

According to Fig. 4, we choose four voltage levels $V_1$, $V_2$, $V_3$, $V_4$ as the voltages corresponding to the four QPSK modulation states. Specifically, $V_1$ represents '00', $V_2$ represents '01', $V_3$ represents '11', and $V_4$ represents '10'. The modulation process of the entire metasurface-based transmitter is shown in Fig. 5 and the key steps are summarized in the following.

*(a) Streaming*: Get the bit stream (010101...) from the information source like pictures or videos;

*(b) Data mapping*: Map the bit stream into the set of QPSK constellation points;

*(c) Sync and pilot mapping*: Map the synchronization sequence and pilots into the set of QPSK constellation points;

*(d) Framing and $\Gamma_n$ mapping*: Form the physical frame, which will be discussed in the next subsection, and obtain the corresponding sequence of $\Gamma_n$;

*(e) Control signal mapping*: According to the actual measured relationship between $\Gamma_n$ and voltage control signal, determine the sequence of voltage control signal for programmable metasurface;

*(f) Controlling*: Control the reflection coefficient of the metasurface according to the obtained sequence of voltage control signal in step (e), and then the reflected electromagnetic wave modulated with the information source messages is transmitted once the incident electromagnetic wave arrives the metasurface.

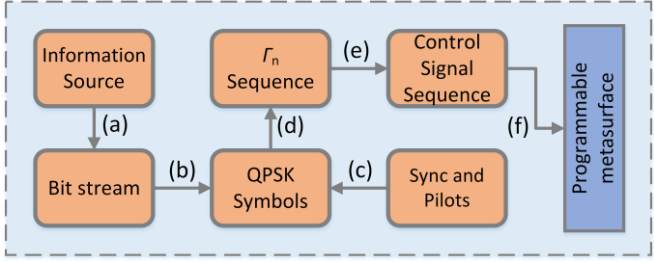

Fig. 5. The modulation process of the transmitter in a metasurface-based single-carrier QPSK wireless communication system.

### C. Frame Structure Design

The proposed frame structure design is shown in Fig. 6. It consists of one synchronization subframe, one pilot subframe and nine data subframes. The synchronization subframe consists of a 420-length synchronization sequence. The pilot subframe consists of 2048 pilot symbols and 160 cyclic prefix (CP) symbols. Similarly, each data subframe also consists of 2048 data symbols and 160 CP symbols. 36864 bits can be transmitted per frame. The transmission rate is mainly determined by the sample rate, i.e., the update rate of the control signal (or equivalently the update rate of the reflection coefficient of the programmable metasurface). We have already achieved 2.048 Mbps real-time transmission rate over the air with 1.25 MSaps sample rate in the proposed single carrier QPSK wireless communication system based on programmable metasurface.

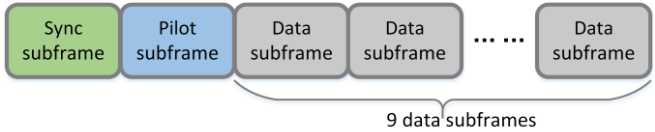

Fig. 6. The proposed frame structure of the metasurface-based single-carrier QPSK wireless communication system.

### D. Receiver Design

The conventional receiver presented in Fig. 3(b) can still be used for the above proposed new transmitter architecture. Such receiver uses a conventional quadrature sampling zero-IF architecture and achieves timing synchronization, carrier synchronization, frequency domain channel estimation, frequency domain channel equalization, and QPSK demodulation by processing the baseband IQ signals as shown in Fig. 7.

The purpose of frame synchronization is to find the beginning of each frame. The metasurface currently used in this paper has a control range of 0 to 255 degrees for the phase modulation of the reflected electromagnetic wave, where the 360 degrees phase modulation coverage has not been achieved yet. Therefore it is not yet possible to implement a conventional synchronization sequence such as Zadoff-Chu (ZC) sequence. Hence we use an extended Barker code sequence to achieve frame synchronization alternatively. Barker code sequence is a binary code group with special rules proposed by R. H. Barker in the early 1950s, and it is the best two-phase sequence which has ideal autocorrelation properties [25]. As Barker code is a kind of binary code, it implies the use of binary phase-shift keying. In the metasurface-based wireless communication system considered in this paper, the Barker code sequence can be readily realized by using the control voltage set of $V_1$ and $V_3$ or $V_2$ and $V_4$, i.e., the change of phase in the carrier wave is 180 degrees. The receiver performs auto-correlation calculation within the search window to achieve frame synchronization by using Barker code.

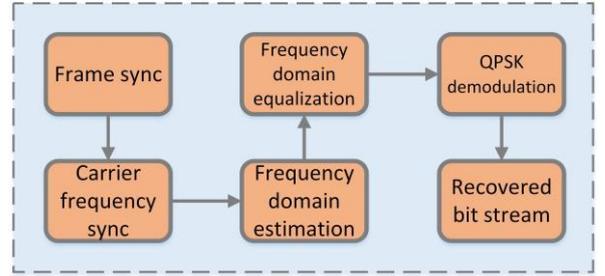

Fig. 7. The demodulation process of the receiver in our metasurface-based single-carrier QPSK wireless communication system.

The receiver employs the CP to perform a joint maximum likelihood estimation of the carrier frequency offset, which is then corrected [26] to eliminate the rotation of the constellation. After synchronization, we implement the CP removal, frequency domain channel estimation (Least Square algorithm), frequency domain channel equalization (Zero Forcing algorithm), QPSK demodulation and bit stream recovery. The corresponding hardware components for our prototype system are presented in Section IV.

## IV. PROTOTYPE SETUP

We present the prototype setup in this section, which illustrates the detailed hardware architecture, including the specification indicator of each hardware module and its role in the prototype system. To implement the programmable metasurface-based single-carrier QPSK wireless communication system described in Section III, we employ the programmable metasurface, control circuit board, several commercial off-the-shelf PXIe modules and software defined radio platforms as follows.

*1) Programmable Metasurface*

The programmable metasurface we designed has already been described in Section II and Section III. It is a reflection-type phase-programmable metasurface with a center frequency of 4 GHz, whose phase profile over the entire surface is controlled by an external input control voltage signal. Fig. 1 shows its schematic and photo. Fig. 4 reveals its tunable characteristics of the phase with the control voltage.

*2) Control Circuit Board*

The control circuit board amplifies the input control votage

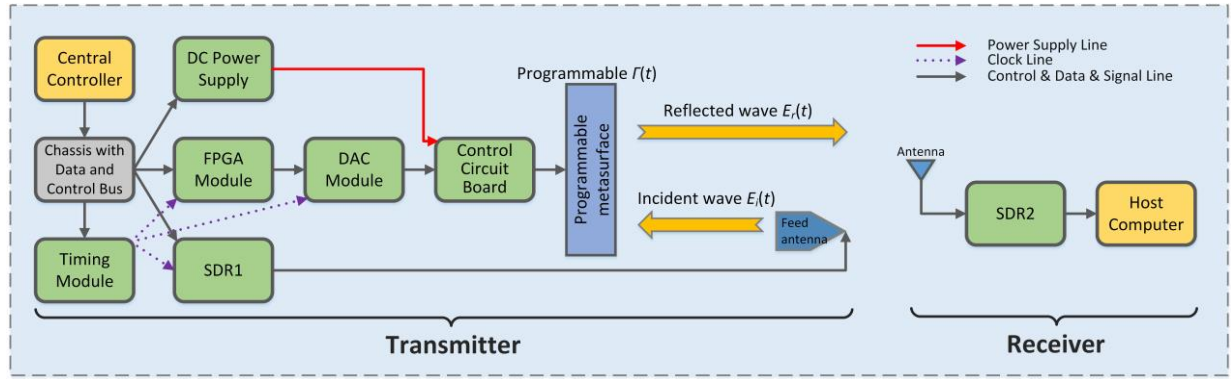

Fig. 8. The detailed hardware architecture of the programmable metasurface-based wireless communication prototype system.

TABLE II

FEATURES OF HARDWARE MODULES

| Module | Name | Features |
|---|---|---|
| Central Controller | PXIe-8135 | Intel Core i7-3610QE quad-core processor 2.3 GHz base frequency CPU |
| Chassis with Data and Control Bus | PXIe-1082 | 8 GB/s bus bandwidth PXIe chassis with 8 slots |
| FPGA Module | PXIe-7966 | Virtex-5 SX95T FPGA, 512 MB DRAM, support peer-to-peer data flow |
| DAC Module | NI-5781 | Analog dual output FlexRIO adapter module with 100MS/s sample rate |
| DC Power Supply | PXI-4110 | Programmable DC power supply with a voltage range of ±20 volts |
| SDR | NI-2943R | 2 RF front ends and 1 Kintex-7 FPGA with carrier frequency from 1.2GHz to 6GHz |
| Timing Module | PXIe-6674T | 10MHz clock based on an onboard precision OCXO reference |

signal to an appropriate value and distributes it to all unit cells of the metasurface.

*3) Central Controller*

The central controller provides the user interface and programming environment for parameter configuration, instrument control and bit file deployment. In addition, the controller reads the local video file to form a bit stream as the information source of the metasurface-based transmitter.

*4) Chassis with Data and Control Bus*

Chassis with data and control bus acts as the interface between the central controller and all PXIe modules, enabling control of all modules and data exchange.

*5) FPGA+DAC Module*

The field programmable gate array (FPGA) and digital-to-analog converter (DAC) module enables the adjustable sampling rate of the metasurface's control digital sequence and converts the digital sequence into analog voltage sequence for subsequent real-time programming of the programmable metasurface.

*6) DC Power Supply*

DC power supply supplies positive and negative 12 volt voltage to the operational amplifier chips on the control board.

*7) SDR*

The software defined radio (SDR) platform provides an integrated hardware and flexible software solution of RF vector signal transceiver. In the metasurface-based transmitter, SDR generates a single tone incident electromagnetic wave to programmable metasurface as the carrier signal. In the receiver, SDR downmixes the received modulated RF signal, and sends the obtained baseband signal to the host computer for synchronization and demodulation processing.

*8) Timing Module*

There is an onboard high precision crystal oscillator on the timing module that provides the same clock source for all modules.

On the basis of the aforementioned description, these hardware components are assembled to implement the prototype of metasurface-based wireless communication shown in Fig. 8. Table II summarizes the corresponding features of selected hardware modules.

The transmitter of our proof-of-concept system is shown in the left part of Fig. 8. It consists of the programmable metasurface, the control circuit board (control signal distribution and amplification) and the PXIe system (control signal sequence generation). We get the source bit stream and realize mapping procedure discussed in Section III on the central controller PXIe-8135 in the PXIe system, and transfer the mapped sequence to the FPGA module PXIe-7966 and DAC module NI-5781 to generate the voltage control signal sequence. The external control circuit board amplifies the control signal to the required voltage range and provides it to all the unit cells of programmable metasurface. Thus the phase modulation of the reflected electromagnetic wave is achieved once the 4 GHz single-tone incident electromagnetic wave is generated from the SDR1.

The receiver in the prototype system is mainly composed of the receiving antenna, the software defined radio platform (SDR2) and the host computer, as shown in the right part of Fig. 8. Baseband IQ data is obtained by the host computer through the conventional RF chain in the SDR2. Synchronization and demodulation are implemented on the host computer in real time.

## V. Experimental evaluation

This section presents the experimental set up to test the proposed programmable metasurface-based single-carrier QPSK wireless communication system in a realistic wireless environment. The main purpose is to demonstrate the feasibility and performance of the developed prototype system in practice. We validate the system's feasibility by visualizing the receiving constellations and video streaming. Furthermore, the BER performance is evaluated by transmitting pseudo-random information bitstream, which is compared with that obtained under the conventional all-SDR-based architecture. In addition, the performance difference between full-activation and half-activation of the programmable metasurface is also discussed.

### A. Experiment Deployment

Real-time video streaming over the air experiment is conducted in a typical indoor environment. The prototype system is shown in Fig. 9 with the main modules labeled, such as the programmable metasurface, PXIe instruments and software defined radio platform. The metasurface-base transmitter is on the right of Fig. 9 and the receiver is on the upper left. The distance between the metasurface and the receiving antenna is 4 meters. The equalized constellation and the source video stream are recovered and displayed as shown on the lower left of Fig. 9.

The main parameters of the implemented prototype system are summarized in Table III. The transmission rate can be further improved by increasing the sampling rate and the modulation order in future work.

TABLE III

Parameters of Metasurface-based Wireless Communication System Here

| Parameter | Value |
|---|---|
| Carrier form | Single carrier |
| Carrier frequency | 4 GHz |
| Modulation method | QPSK |
| Sampling rate | 1.25 MSps |
| Frame size | 22500 samples |
| Transmission rate | 2.048 Mbps |

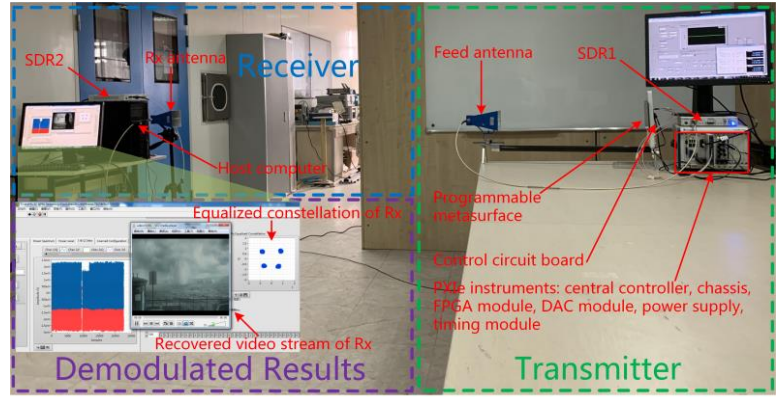

Fig. 9. The prototype of the proposed programmable metasurface-based single-carrier QPSK wireless communication system.

### B. Measurement Results

A series of measurements are carried out in this experiment. Experiment results show that the QPSK constellation diagram after equalization is clear and stable, and the video stream can be transmitted smoothly and clearly even in the absence of channel coding. This strongly demonstrates the feasibility of the proposed metasurface-based communication system. The measured constellation diagrams under different transmission power are shown in Fig. 10. It is observed that the higher the transmission power is, the denser the constellation points are, which indicates the improved BER performance, as expected. Furthermore, it is observed that the distribution of the constellation points is not square, which is expected since the amplitude response of the metasurface used is non-uniform under different phase responses. The programmable metasurface we designed has not yet achieved the same reflection gain for different phases, which will be further improved in the future work. However, it is observed that the prototype system with unevenly distributed constellation points can still work reliably.

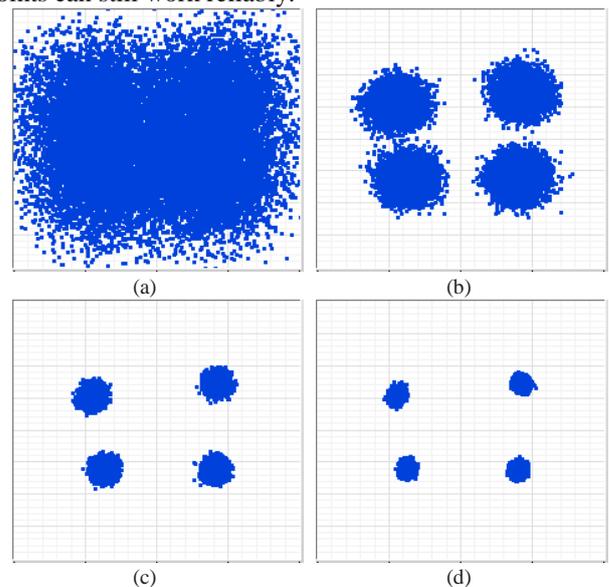

Fig. 10. The measured constellation diagrams under different transmission power: (a) -55 dBm (b) -45 dBm (c) -35 dBm (d) -25 dBm

In addition, we also design a comparative experiment to compare the BER performance between the proposed programmable metasurface-based prototype system and the conventional all-SDR-based wireless communication system. Both systems are based on the same hardware architecture shown in Fig. 8. However, for the metasurface-based wireless communication system, SDR1 only provides a single-tone carrier signal and the programmable metasurface implements QPSK modulation of electromagnetic wave. By contrast, the conventional all-SDR-based system is programmed such that SDR1 constantly sends the same QPSK frame structure defined in Section III over the air, using the same frequency, sampling rate and other parameters as the metasurface-based transmitter, while providing a fixed control voltage to programmable metasurface. In other words, for the all-SDR-based system, the metasurface does not modulate the incident electromagnetic wave but only acts as a reflector. The receiver design of the two systems are exactly the same.

By varying the transmit power of SDR1, the BER performance of the two systems are measured. For each transmit power level, $10^4$ frames are transmitted over the air, which contains a total of $3.6864\times10^2$ Mbits. Since the transmitted pseudo-random bit information is fixed, the receiver knows the correct bit information a prior, based on which the BER at the receiving end is calculated.

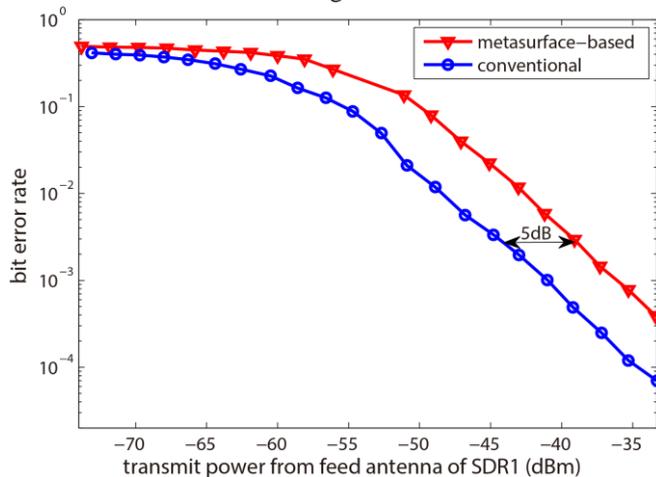

Fig. 11. The BER performance of programmable metasurface-based single-carrier QPSK wireless communication system versus conventional all-SDR-based single-carrier QPSK wireless communication system.

Fig. 11 plots the BER performance of the programmable metasurface-based and conventional all-SDR-based QPSK wireless communication systems. It is observed that with a slight increase of the transmit power by 5 dB, the metasurface-based system is able to achieve the same BER performance as the conventional one. Such a performance gap is expected because the sampling function used in our current metasurface-based transmitter is a rectangular window function, which will cause the energy of the reflected electromagnetic wave to leak out of the effective frequency band, resulting in loss of signal energy. In addition, the imperfection of the feed network results in the control voltages obtained by each unit cell of metasurface not being synchronized ideally, and the op-amp circuit of the control board also brings additional noise. These aspects will be improved and optimized in future work. However, such measurement results are encouraging, considering that metasurfaces are simpler in structure, more cost-effective, easier to implement for large-scale channels than the conventional systems. Furthermore, the working frequency band of metasurfaces may span from microwave to visible light, which may expand the application prospect of metasurface greatly.

We also compare the cases when all unit cells of the programmable metasurface are activated and only half (left half or right half) are activated. When only half of the unit cells are activated, the control voltage of the activated unit cells changes periodically to perform QPSK modulation. Meanwhile the control voltage of the other half of the unit cells that are not activated remains unchanged. As Fig. 12(a) shows, the measured SNR with full-activation configuration is about 6 dB higher than the half-activation configuration in the same incident carrier signal power from the feed antenna, which indicates that the metasurface-based transmitter's array gain is related to the number of activated unit cells and their total aperture size. When the programmable metasurface is in the half-activation state, the number of unit cells performing QPSK modulation is half of that in full-activation state. At the first glance, it may expect that the total effective aperture should be reduced by half, and the corresponding loss of array gain should be 3 dB instead of 6 dB. However, in practice, the loss of array gain is also affected by the other half inactive unit cells, which do not perform QPSK modulation but reflect the incident electromagnetic waves back directly. These reflected electromagnetic waves are superimposed in space with the other half of the reflected electromagnetic waves containing phase modulation information. The superposition of these two electromagnetic waves at the receiving antenna causes the cancellation of the received signals and leads to further loss of the equivalent aperture of programmable metasurface. The BER performance is also measured and shown in Fig. 12(b). It is observed that the BER performance in a full-activation configuration is better than the half-activaion configuration. Fig. 12(b) also shows that the same BER performance as in the full-activation state can be achieved by increasing the transmission power by 6 dB in the half-activation state. The BER performance corresponding to the left half-activation configuration and the right half-activation configuration are almost identical, which is expected since the receiving antenna is placed in the straight forward direction of metasurface and its two half sides are spatially symmetric.

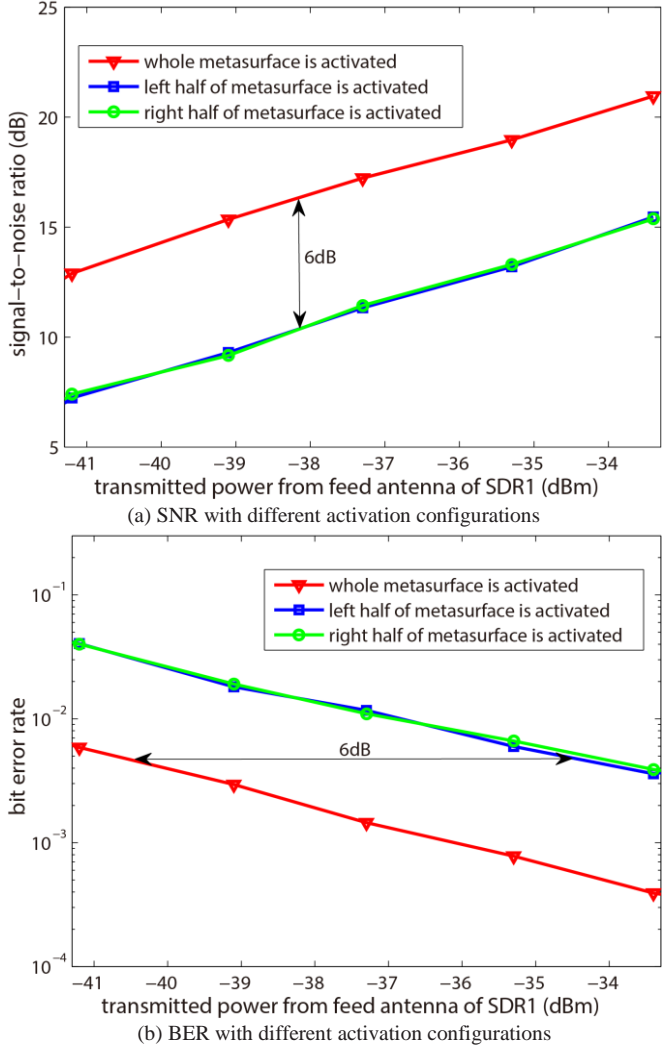

Fig. 12. SNR and BER comparison with all unit cells of the programmable metasurface are activated and only half (left half or right half) are activated. (a) SNR curves with different activation configurations. (b) BER curves with different activation configurations.

## VI. DISCUSSION AND FUTURE WORK

In this paper, experiments and measurement results have demonstrated the great potential of using programmable metasurface as transmitter in wireless communication systems. Some important performance characteristics have been revealed. The hardware architecture of the metasurface-based transmitter in this paper does not require any filter, wideband mixer or power amplifier, rendering it an attractive technology for realizing cost-effective wireless communications. Furthermore, the proposed metasurface-based architecture has great potential to generate multi-beams and complex radio signals, since in principle, each unit cell of the programmable metasurface can be controlled independently. The proposed architecture has great potential for a wide range of applications, from microwave to optical frequencies, and provides a new way to resolve the issue of hardware constraints in future advanced wireless communication systems, such as massive MIMO millimeter wave communication and artificial intelligence embedded systems. The exploitation of programmable metasurface for wireless communication systems presents a new research field that is still in its early stage. While our work in this paper shows some promising results, there are still many challenges worth further exploration in future research.

In the following, we outline some important open research topics for metasurface-based wireless communication systems, including theoretical modeling, metasurface-based receiver, high-order transmission mechanism design and coverage enhancement.

*1) Theoretical modeling*: As the architecture of metasurface-based transmitter is significantly different from the conventional ones, it is of great importance to develop accurate analytical signal models for it. The nonlinearity of phase response and the effect of charge/discharge of varactor diodes should be taken into account in the theoretical modeling as the hardware non-ideal characteristics of the programmable metasurface. How to model these different hardware characteristics is the key to study the issues of channel capacity, energy efficiency, and transmission mechanism of metasurface-based communications.

*2) Metasurface-based receiver*: Exploiting programmable metasurface in the receiver is promising for the purpose of achieving enhanced performance. Particularly, programmable metasurface could be reconceptualized as programmable reflective antenna array or programmable beam antenna array, which can be then applied in metasurface-based receiver design, such as channel estimation, hybrid beamforming and interference control. Furthermore, by integrating programmable metasurface into the receiver and obtaining the integrated architecture of the transceiver based on programmable metasurface, it is likely to inherit features such as channel reciprocity in time-division duplexing (TDD) wireless communication systems.

*3) High-order modulation and waveform design*: High-order modulation and advanced waveform design can greatly improve spectrum utilization. In the future, on one hand, we may start with conventional high-order modulation schemes and waveform designs such as quadrature amplitude modulation (QAM), discrete multitone modulation (DMT) and orthogonal frequency division multiplexing (OFDM) to explore transmission techniques that are suitable for programmable metasurface. On the other hand, machine learning (ML), especially deep learning (DL), can be applied to achieve appropriate modulation and waveform schemes for metasurface-based systems, considering the nonlinear characteristics of metasurface.

*4) Beam steering and coverage enhancement*: Programmable metasurface can alter the transmission path of electromagnetic waves, and thus can be used for beam steering and coverage enhancement. For millimeter communication in particular, programmable metasurfaces arranged in the wireless channels can be used to transform some non line of sight (NLOS) channels into line of sight (LOS) channels to improve coverage performance. The coding method of programmable metasurface for electromagnetic wave manipulation is worthy for further studies, including wavefront controlling and polarization direction regulation, to realize beam steering and beam tracking based on programmable metasurface in the future.

## VII. CONCLUSION

In this paper, we have presented the use of programmable metasurface as low-cost transmitter for wireless communications. The basic principle and method of designing such metasurface-based system have been introduced. We successfully demonstrated a programmable metasurface with 8×16 tunable unit cells for a QPSK wireless communication prototype system, which validated the feasibility of the proposed metasurface-based system architecture. The experimental results demonstrated that the metasurface-based architecture is able to achieve comparable performance as the conventional architecture, but with less hardware complexity and thus leading to a promising new architecture for wireless communications.